\documentclass[preprints,article,accept,moreauthors,pdftex]{Definitions/mdpi} 

\firstpage{1} 
\pubvolume{1}
\issuenum{1}
\articlenumber{0}
\pubyear{2021}
\copyrightyear{2020}
\datereceived{} 
\dateaccepted{} 
\datepublished{} 
\hreflink{https://doi.org/} 
\pdfoutput=1

\usepackage{amssymb,amsfonts}
\usepackage{psfrag}
\usepackage{pstool}
\usepackage{mathtools}
\usepackage{ulem}
\usepackage{pifont}

\newcommand{\B}{\mathbf}
\newcommand{\BO}{\B{B}_0}
\newcommand{\EO}{\B{E}_0}
\newcommand{\HO}{\B{H}_0}
\newcommand{\DO}{\B{D}_0}
\newcommand{\tx}{\text}
\newcommand{\oo}[1]{\overline{\overline{#1}}}
\newcommand{\X}{\B{\hat{x}}}
\newcommand{\Z}{\B{\hat{z}}}
\newcommand{\K}{\B{k}}
\newcommand{\s}{\B{S}}

\newcommand{\sgn}{\operatorname{sgn}}

\Title{Electromagnetic Wave Scattering from a Moving Medium with Stationary Interface across the Interluminal Regime}

\TitleCitation{Title}

\Author{Zo\'{e}-Lise Deck-L\'eger$^{1}$, Xuezhi Zheng$^2$ and Christophe Caloz$^{2}$* }

\AuthorNames{Zo\'{e}-Lise Deck-L\'eger, Xuezhi Zheng and Christophe Caloz}

\AuthorCitation{Deck-L\'eger, Z.-L.; Zheng, X.; Caloz, C.}

\address{%
$^{1}$ \quad Department of Electrical Engineering, Polytechnique Montr\'eal, 2500, ch. de Polytechnique, Montr\'eal, Montreal, H3T 1J4, Canada; zoe-lise.deck-leger@polymtl.ca\\
$^{2}$ \quad Faculty of Engineering Science, KU Leuven, P.O. Box 2444, 3001 Leuven, Belgium; xuezhi.zheng@kuleuven.be; christophe.caloz@kuleuven.be}

\corres{To whom correspondence should be addressed}

\abstract{This paper extends current knowledge on electromagnetic wave scattering from bounded moving media in several regards. First, it complements the usual dispersion relation of moving media, $\omega(\theta_\mathbf{k})$ ($\theta_\mathbf{k}$: phase velocity direction, associated with the wave vector, $\mathbf{k}$), with the equally important impedance relation, $\eta(\theta_\mathbf{S})$ ($\theta_\mathbf{S}$: group velocity direction, associated with the Poynting vector, $\mathbf{S}$). Second, it explains the interluminal-regime phenomenon of double-downstream wave transmission across a stationary interface between a regular medium and the moving medium, assuming motion perpendicular to the interface, and shows that the related waves are symmetric in terms of the energy refraction angle, while being asymmetric in terms of the phase refraction angle, with one of the waves subject to negative refraction, and shows that the wave impedances of the two transmitted waves are equal. Third, it generalizes the problem to the case where the medium moves obliquely with respect to the interface. Finally, it highlights the connection between this problem and a spacetime modulated medium. }


\begin{document}
\section{Introduction}
\label{sec:introduction}

The problem of light radiation and scattering from moving media has been abundantly studied. The first studies pertained to radiation from stars and led to the discoveries of related frequency shifts by Doppler~\cite{Dopplerbook} and velocity aberrations by Bradley~\cite{bradley1729}. Then Einstein generalized the Doppler-shift, Bradely-aberration and Fresnel-reflection formulas to relativistic velocities~\cite{einstein1905}. Next Minkowski, leveraging the tools of relativity, derived electromagnetic equations describing light scattering phenomena observed in moving media, in particular the transformation of isotropic media, such as simple dielectrics, into `bianisotropic' media~\cite{kong1972} when set into motion~\cite{minkowski1908}. Upon this basis, many investigations on scattering in and from moving dielectrics have been reported since the early twentieth century, e.g.,~\cite{yeh1965,tsai1967,bolotovski1989}.

The vast majority of the research on moving media has concerned objects that move as a whole, with their edges co-moving with the medium. But moving media can also be bounded by stationary interfaces. Such a situation is very common when the motion is parallel to the interfaces, as is the case for instance of the water flow of a river parallel to its banks. Wave scattering from this type of interface involves effects such as velocity-dependent positive or negative refraction~\cite{grzergorczyk2006} and modified Goos-H\"{a}nchen shifting~\cite{wang2012}. Less common is the situation where the motion of the medium is perpendicular to a stationary interface, such as is the case for
a conveyor belt with fixed ends~\cite{vehmas2014}. This type of boundary, although of even greater interest, as shall be shown, has been much less studied. The corresponding scattering coefficients were derived for the subluminal regime in~\cite{censor1969} and the interluminal-regime phenomenon of double-wave transmission was described for the case of normal incidence in~\cite{vehmas2014}, but much of the related physics remains to be unveiled.

Here, we extend the established knowledge on moving media with stationary boundaries. We first recall the bianisotropic nature and properties of unbounded moving media. Upon this basis, we then re-derive their isofrequency relation and establish their impedance relation. We next study the properties of waves scattered from such media when bounded by stationary interfaces. Specifically, we describe the deflection angle of the phase and group velocities for the whole range of medium velocities, from $-c$ to $c$, and calculate the corresponding wave impedance, for medium motion perpendicular to the interface. Moreover, we graphically solve velocities and the impedance of waves scattered at the interface when the medium motion is at an oblique angle with respect to the interface and show that this configuration can involve negative refraction. Finally, we provide a connection between the studied moving medium problem and the related problem of a spacetime-modulated medium.

\section{Description of the Problem}\label{sec:description}

\begin{figure}[ht] \includegraphics[width=1\columnwidth]{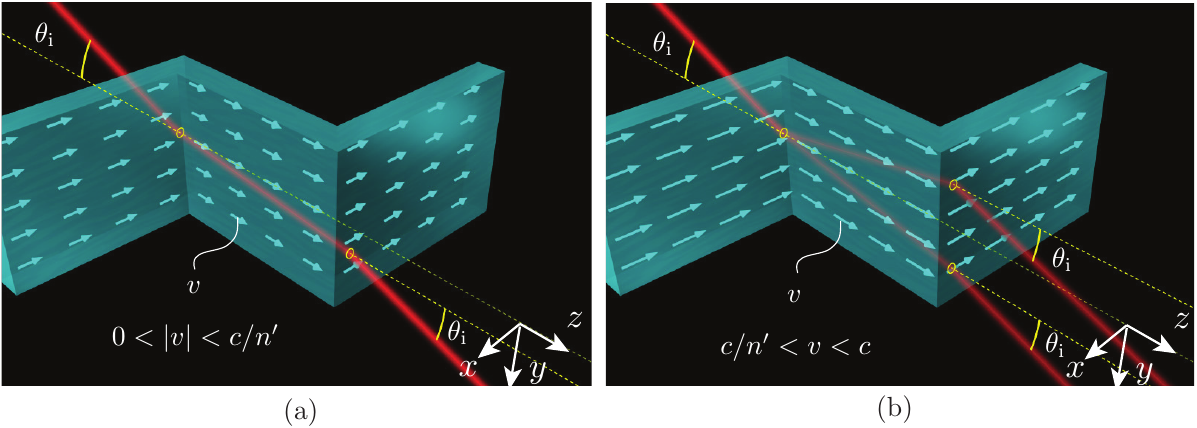}
	\caption{Wave scattering at a stationary interface between a simple (isotropic, dispersionless and linear) medium, such as air or a standard dielectric fluid, and a moving medium with uniform motion of velocity $v$ perpendicular to the interface and rest refractive index (refractive index if the medium were at rest) $n'$. (a)~Subluminal regime ($|v|\le c/n'$). (b)~Interluminal regime ($|v|\ge c/n'$).}
	\label{fig:Fizeau}
\end{figure}

Figure~\ref{fig:Fizeau} depicts the problem of interest, namely the scattering of electromagnetic waves at a stationary interface between a simple stationary medium, such as air or a simple dielectric, and a moving medium. The moving medium is here assumed to be a fluid flowing at a uniform velocity through a double-bend structure, so that the interfaces at the the bends are stationary with the medium moving perpendicularly to them. Electromagnetic waves are obliquely incident at the interface formed by one of the bends, refract into the moving fluid, and exit the fluid at the other interface. The setup of Fig.~1 is inspired by the Fizeau experiment~\cite{fizeau1851hypotheses}\footnote{In this experiment, Fizeau measured the speed of light in moving fluid and hence confirmed the 1818 prediction of Fresnel~\cite{fresnel1818} that light was dragged by a moving medium. In contrast to the case of Fig.~1, his experiment, using circular hoses as opposed to elongated containers (as here) to support the fluid, was restricted to normal incidence and ignored the effect of scattering at the interface between the incoming-wave air region and the fluid medium.}, but it is to be mostly considered as a thought experiment insofar as the some of the fluid velocities of interest would be difficult to attain in practice.

The medium is assumed to be moving at the constant and uniform velocity $v$ with respect to the laboratory frame, and we use the relativity convention of the unprimed and primed variables to denote quantities that are measured in the laboratory frame and in the moving-medium frame, respectively. We distinguish the following two regimes:

\begin{equation}\label{eq:two_regimes}
\begin{cases}
\begin{rcases}
|v|\in[0,v_\text{w}']:\:\text{subluminal regime} \\
|v|\in[v_\text{w}',c]:\:\text{interluminal regime}
\end{rcases}
\end{cases}\hspace{-3mm},\quad
\text{with}\quad~v_\text{w}'=\frac{c}{n'}
\end{equation}
being the wave velocity measured in the frame of the moving medium, where the fluid is at rest and has the refractive index $n'$. According to the definition~\eqref{eq:two_regimes}, the subluminal regime corresponds to medium velocities that are smaller than both the wave velocity in the medium at rest ($v_\text{w}'$) and the wave velocity in free space ($c$), while the interluminal regime corresponds to medium velocities that are still smaller than the wave velocity in free space -- the maximal velocity for matter -- but larger than the wave velocity in the medium at rest\footnote{In the case of a moving charged particle, as opposed to that of a moving medium considered here, the interluminal regime corresponds to the Cherenkov regime, where the particle radiates energy because it moves faster than the wave velocity in the medium (e.g., heavy water)~\cite{cherenkov1936,tamm1937}.}.

Figures~\ref{fig:Fizeau}(a) and~\ref{fig:Fizeau}(b) describe the scattering phenomenology of the system in the subluminal and interluminal regimes, respectively. As will be shown in Sec.~\ref{sec:scatter}, a single wave is transmitted in the moving medium in the subluminal regime, while two waves, with symmetric group velocities, are transmitted in the moving medium in the interluminal regime.

\section{Moving-Medium Bianisotropy}\label{eq:const_rel}

Assuming time-harmonic plane waves of the form
\begin{equation}\label{eq:th_fields}
\B{A}=\B{A}_0e^{i(\K\cdot\B{r}-\omega t)},
\end{equation}
where $\B{A}=\B{E}, \, \B{H},\, \B{D},\, \B{B}$ are the usual electromagnetic fields, and $\mathbf{k}$ and $\omega$ are the spatial frequency (or wave vector) and temporal frequency, respectively, the Maxwell equations in a source-free region of space take the form
\begin{equation}\label{eq:max}
\K\times \EO=\omega\BO, \quad \K\times \HO=-\omega\DO, \quad \K\cdot\BO=0, \quad \K\cdot\DO=0.
\end{equation}
In the frame of the moving medium, the constitutive relations are

\begin{equation}\label{eq:D'}
\DO'=\epsilon' \EO', \qquad \BO'=\mu' \HO',
\end{equation}
\noindent where $\epsilon'$ and $\mu'$ are scalar constants under the assumption that the medium at rest is isotropic, dispersionless and linear. The corresponding relations in the frame of the laboratory are found by applying the Lorentz transformations~\cite{Lorentz1937} to the fields in~\eqref{eq:D'} and rearranging the result so as to express $\B D$ and $\B B$ in terms of $\B E$ and $\B H$. This leads to the bianisotropic constitutive relations~\cite{kong1972,kong2005}
\begin{subequations}\label{eq:const}
\begin{equation}\label{eq:const_D}
\DO=\oo{\epsilon}\cdot \EO+\oo\xi\cdot\HO,
\end{equation}
\begin{equation}\label{eq:const_B}
\BO=\oo{\mu} \cdot\HO+\oo\zeta\cdot\EO,
\end{equation}
\end{subequations}
where
\begin{subequations}\label{eq:eqbianistens}
\begin{equation}\label{eq:eqbianistensa}
\oo\epsilon=
\epsilon'\begin{bmatrix}
\alpha & 0 & 0\\
0 & \alpha & 0\\
0 & 0 & 1
\end{bmatrix}, \quad
\oo\mu=
\mu'\begin{bmatrix}
\alpha & 0 & 0\\
0 & \alpha & 0\\
0 & 0 & 1
\end{bmatrix}, \quad
\oo\xi=
\begin{bmatrix}
0 & \chi/c & 0\\
-\chi/c & 0 & 0\\
0& 0 & 0
\end{bmatrix}, \quad \oo\zeta=\oo\xi^T,
\end{equation}
with
\begin{equation}\label{eq:alpha_chi}
\alpha=\frac{1-\beta^2}{1-\beta^2{n'}^2}, \qquad \chi=\beta\frac{1-{n'}^2}{1-\beta^2{n'}^2},
\end{equation}
\end{subequations}
where $\beta=v/c$. In these relations, the signs of $\alpha$ and $\chi$ are opposite in the subluminal and the interluminal regime. This change in sign leads to drastically different physics in the two regimes, as we shall see. The parameters $\alpha$ and $\chi$ are relativistic ($v/c$-dependent), with $\chi$ being in addition time-reversal asymmetric, i.e., $\chi(-v)=-\chi(v)$, and thus inducing nonreciprocity, also manifested by the violation of the reciprocity condition $\oo{\zeta}=-\oo{\xi}^T$~\cite{caloz2018}.

According to the last two equations in~\eqref{eq:max}, we have $\K\perp\DO$ and $\K\perp\BO$, so that $\K\|(\DO\times\BO)$. In contrast, $\EO$ and $\HO$ are generally not parallel to $\DO$ and $\BO$, respectively, and therefore $\B{S}_0=\EO\times\HO$ is not parallel to $\K$, in a bianisotropic medium. The angle of $\K$ with respect to the direction of motion, or the phase angle, is defined as $\theta_\K=\arctan(k_x/k_z)$, while the angle of $\B{S}_0$ with respect to the direction of motion, or the energy angle, must be computed separately for the p- and s-polarizations.

Applying the identity $\K\times\B{A}=\oo\K\cdot\B{A}$ with 
\begin{equation}\label{eq:Ktens}
\oo\K=\begin{bmatrix}
                           0 & -k_z & k_y \\
                           k_z & 0 & -k_x \\
                           -k_y & k_x & 0
                         \end{bmatrix},
\end{equation}
\noindent to the first two equations in~\eqref{eq:max} and substituting the constitutive relations~\eqref{eq:const} in the resulting equations leads to the dyadic relations

\begin{equation}
\oo\K\cdot \B{E_0}=\omega\left(\oo{\mu}\cdot \B H_0+\oo\zeta\cdot\B E_0\right), \quad \oo\K\cdot \B{H}_0=-\omega\left(\oo{\epsilon}\cdot \B E_0+\oo\xi\cdot\B H_0\right),
\end{equation}
and solving these equations for $\B{H}_0$ and $\B{E}_0$ yields
\begin{equation}
\B{H}_0=\frac{1}{\omega}\oo\mu^{-1}\cdot\left(\oo\K-\omega\oo\zeta\right)\cdot\B E_0, \quad \B{E}_0=-\frac{1}{\omega}\oo\epsilon^{-1}\cdot\left(\oo\K+\omega\oo\xi\right)\cdot\B H_0.
\end{equation}
Assuming then, without loss of generality, that scattering occurs in the $xz$-plane, so that $k_y=0$ in~\eqref{eq:Ktens}, these relations become, upon substitution of the so-reduced tensor $\oo\K$ and of the constitutive tensors in~\eqref{eq:eqbianistens},
\begin{subequations}\label{eq:max_explicit}
\begin{equation}\label{eq:h0}
\B{H}_0=\frac{1}{\omega\mu'\alpha}\begin{bmatrix}
  0 & -(k_z-\omega\chi/c) & 0 \\
  k_z-\omega\chi/c & 0 & -k_x \\
  0 & \alpha k_x & 0
\end{bmatrix}\B{E}_0
\end{equation}
\begin{equation}\label{eq:e0}
\B{E}_0=-\frac{1}{\omega\epsilon'\alpha}\begin{bmatrix}
  0 & -(k_z-\omega\chi/c) & 0 \\
 k_z-\omega\chi/c & 0 & -k_x \\
  0 & \alpha k_x & 0
\end{bmatrix}\B{H}_0.
\end{equation}
\end{subequations}

Since the medium is achiral, according to the expression of $\oo\xi$ in~\eqref{eq:eqbianistensa}~\cite{caolz2020_1,caolz2020_2}, it does not induce any polarization rotation, and the problem decouples therefore into p- and s-polarized waves. For p-polarized waves, we have $\B{H}_0=H_{0y}\hat{\B{y}}$, and we find from~\eqref{eq:e0} the $\EO$ field components
\begin{subequations}\label{eq:p_max}
\begin{equation}\label{eq:e0xz}
 E_{0x}=\frac{k_z-\omega\chi/c}{\omega\epsilon'\alpha}H_{0y},\quad E_{0z}=-\frac{k_x}{\omega\epsilon'}H_{0y}.
\end{equation}
For later use, we also note from~\eqref{eq:h0} that
\begin{equation}\label{eq:hoy}
H_{0y}=\frac{1}{\omega\mu'\alpha}\left[\left(k_z-\omega\chi/c\right)E_{0x} -k_x E_{0z}\right].
\end{equation}
\end{subequations}

Inserting relations~\eqref{eq:e0xz} into the expression of the Poynting vector $\B{S}_0=-(E_{0z}\hat{\B{x}}-E_{0x}\hat{\B{z}})H_{0y}$ yields then the angle of this vector with respect to the medium motion direction ($z$) as
\begin{equation}\label{eq:s_angles}
\theta_{\s}=\arctan\left(\frac{-E_{0z}}{E_{0x}}\right)=\arctan\left(\frac{\alpha k_{x}}{k_{z}-\omega\chi/c}\right).
\end{equation}

For s-polarized waves, we have $\B{E}_0=E_{0y}\hat{\B{y}}$, and we find  from~\eqref{eq:h0} the $\HO$ components
\begin{subequations}\label{eq:s_max}
\begin{equation}\label{eq:h0xz}
 H_{0x}=-\frac{k_z-\omega\chi/c}{\omega\mu'\alpha}E_{0y},\quad H_{0z}=\frac{k_x}{\omega\mu'}E_{0y},
\end{equation}
and, again for later use, we  note from~\eqref{eq:e0} that
\begin{equation}\label{eq:eoy}
E_{0y}=\frac{-1}{\omega\epsilon'\alpha}\left[\left(k_z-\omega\chi/c\right)H_{0x} -k_x H_{0z}\right].
\end{equation}
\end{subequations}

Inserting relations~\eqref{eq:h0xz} into the expression of the Poynting vector, which is now $\B{S}_0=(H_{0z}\hat{\B{x}}-H_{0x}\hat{\B{z}})E_{0y}$, yields then the same Poynting angle as that of the p-polarized wave, given by the second equality of~\eqref{eq:s_angles}.

\begin{figure}[ht] \includegraphics[width=1\columnwidth]{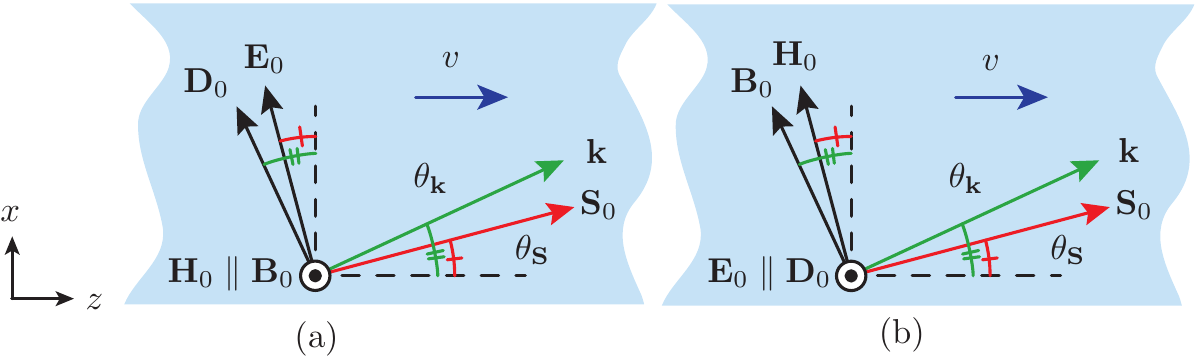}
	\caption{Electromagnetic vector configurations in the moving medium. (a)~p-polarization. (b)~s-polarization.}
	\label{fig:waves}
\end{figure}

Figure~\ref{fig:waves} shows the vector configurations for two wave polarizations. For the p-polarization, there is a splitting of the $\DO$ and $\EO$ fields, while for the s-polarization, the splitting is between the $\BO$ and $\HO$ fields. 

\section{Dispersion Relation and Isofrequency Diagram}\label{sec:refr_ind_vel}
The dispersion relation is obtained by eliminating the field quantities in the set of equations~\eqref{eq:max_explicit} (inserting~\eqref{eq:e0} into~\eqref{eq:h0}), and setting the determinant of the resulting system to zero, which yields~\cite{kong2005}
\begin{equation}\label{eq:disp_0}
\left(k_z-\frac{\omega}{c}\chi\right)^2+\alpha k_x^2=\left(\alpha n'\frac{\omega}{c}\right)^2.
\end{equation}
This may alternatively be written in polar form, upon substituting $k_z=k\cos\theta_\K$ and $k_x=k\sin\theta_\K$ with $k=\sqrt{k_z^2+k_x^2}=n(\theta_\K) \omega/c$ in this relation, as
	\begin{equation}\label{eq:neff} \left(n(\theta_\K)\cos\theta_\K-\chi\right)^2+\alpha\left(n(\theta_\K)\sin\theta_\K\right)^2=\left(\alpha n'\right)^2,
	\end{equation}
where $n$, the refractive index of the medium as seen in the laboratory frame, is a function of $\theta_\B{k}$ and should not be confused with $n'$, the refractive index seen in the moving frame. The independence of the refractive index from $\omega$ in \eqref{eq:neff} indicates that the medium remains temporally nondispersive ($n\neq{n}(\omega)$) when it is set in motion. In contrast, the dependence of the refractive index on the propagation angle ($n=n(\theta_\B{k})$) indicates that the velocity of the medium induces anisotropy. This is due to the drag of the medium on the wave. The anisotropy is elliptical for $\alpha>0$ (subluminal regime) or hyperbolic for $\alpha<0$ (interluminal regime), and the relation is off-centered, with the center at $(k_z,k_x)=(\chi \omega/c,0)$\footnote{The rest-medium relations are retrieved by setting $v=0$. We obtain then $\alpha=1$ and $\chi=0$ from~\eqref{eq:alpha_chi}, and hence circular refractive index curves with radius~$n'\omega/c$.}.

We then compute, from the dispersion relation~\eqref{eq:disp_0}, the phase and group velocities. The phase velocity vector $\B{v}_\tx{p}$ is found by solving $n(\theta_\K)$ in~\eqref{eq:disp_0}, which yields~\cite{bolotovskii1974} 
\begin{equation}\label{eq:vp} \B{v}_\tx{p}=\frac{c}{n(\theta_\K)}\mathbf{\hat{k}}=\frac{c\left(\cos^2\theta_\K+\alpha\sin^2\theta_\K\right)}{\chi\cos(\theta_\K)\pm\sqrt{\chi^2\cos^2\theta_\K-(\cos^2\theta_\K+\alpha\sin^2\theta_\K)(\chi^2-\alpha^2{n'}^2)}}\hat\K,
\end{equation}
where $\mathbf{\hat{k}}=\mathbf{k}/k$. The phase velocity is hence directed along the radial vector $\mathbf{\hat{k}}$, subtended by the polar angle $\theta_\mathbf{k}$. The group velocity vector is found by differentiating~\eqref{eq:disp_0} with respect to $k_x$ and $k_z$, and grouping the two results, which leads to~\cite{bolotovskii1974}
\begin{equation}\label{eq:vg}
	\B{v}_\tx{g}=\nabla_\K\omega(\mathbf{k})=\frac{\alpha k_x \X+(k_z-\omega\chi/c)\Z}{\omega (\alpha^2 {n'}^2/c^2)+\chi/c(k_z-\omega\chi/c)}.
	\end{equation}
It can be shown that the phase velocity does not correspond to the velocity addition rule derived by Einstein, whereas the group velocity does~\cite{pauli1958}.

The angles for the the phase and group velocities are found from the phase and group velocity expressions~\eqref{eq:vp} and~\eqref{eq:vg} as
\begin{subequations}\label{eq:angles_pg}
\begin{equation}\label{eq:thetak}
\theta_{\K}=\arctan\left(\frac{k_{x}}{k_{z}}\right),
\end{equation}
\begin{equation}\label{eq:thetas}
\theta_{\s}=\arctan\left(\frac{\alpha k_{x}}{k_{z}-\omega\chi/c}\right).
\end{equation}
\end{subequations}
The group velocity vector $\B{v}_\tx{g}$, as a gradient, is perpendicular to the frequency contour curves $\omega(k_z,k_x)$ evaluated at a given frequency $\omega=\omega_0$. The Poynting vector is also perpendicular to the contour curves~\cite{kong2005}, and thus the vectors $\B{v}_\tx{g}$ and $\B{S}$ share the same polar angle, $\theta_\s$. For the specific case of moving media, this is verified by comparing the second equation of~\eqref{eq:angles_pg} with the angle of the Poynting vector calculated in~\eqref{eq:s_angles}.

\begin{figure}[ht] \includegraphics[width=1\columnwidth]{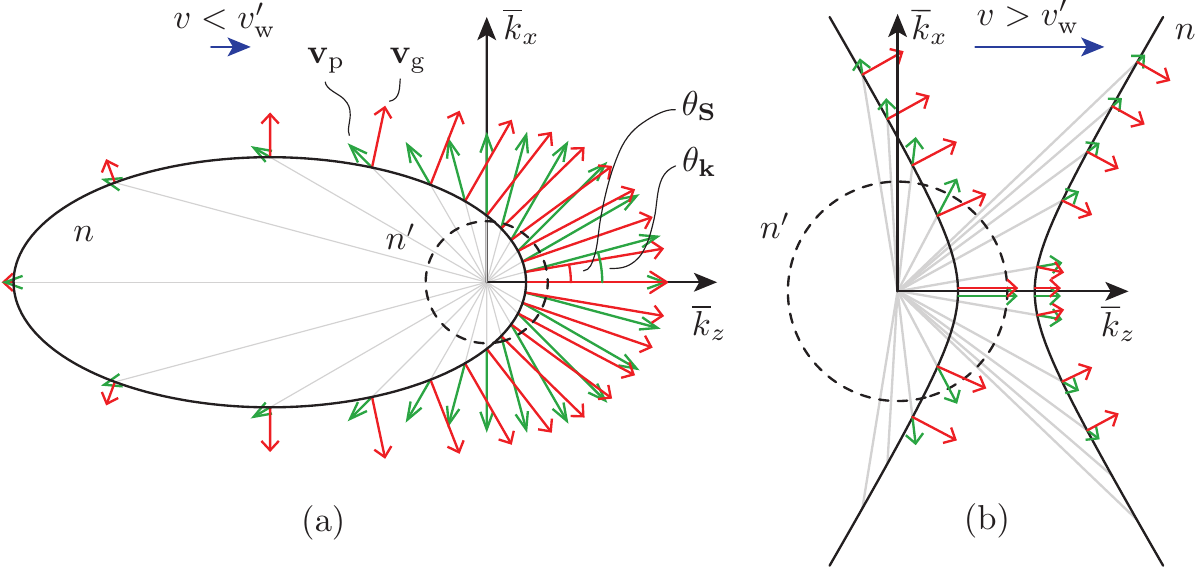}
	\caption{Dispersion relation, with phase and group velocity vectors for a moving dielectric (fluid) medium with refractive index $n'=2$ (wave velocity in the rest medium: $v_\text{w}'=c/2$). (a)~Subluminal regime, with velocity $v=0.9v_\text{w}'<v_\text{w}'$. (b)~Interluminal regime, with $v=1.6v_\text{w}'>v_\text{w}'$. The dashed circles correspond to the isotropic curves at $v=0$.}
	\label{fig:refractive}
\end{figure}

Figure~\ref{fig:refractive} plots the dispersion relations~\eqref{eq:disp_0}, with $\bar{k}_{x,z}=k_{x,z}/k_0$, and the phase and group velocity vectors~\eqref{eq:vp} and~\eqref{eq:vg}. Figure~\ref{fig:refractive}(a) corresponds to the subluminal regime and Fig.~\ref{fig:refractive}(b) corresponds to the interluminal regime.
In the subluminal regime (Fig.~\ref{fig:refractive}(a)), the velocity is less for purely upstream propagation, $\theta_\B{k}=\pi$, than for purely downstream propagation, $\theta_\B{k}=0$\footnote{The exact wave velocity for propagation parallel to the motion was found by Einstein as $v_\text{w}=[(c/n')\pm{v}]/[1+(v/n')/c]$, which reduces to the Fresnel-Fizeau result when $v\ll{c}$, using $1/(1+x)\approx{1-x}$ for $x\ll{1}$.}. When the phase velocity is perpendicular to the motion, the group velocity is deflected by the stream, a result of the drag.
In the interluminal regime (Fig~\ref{fig:refractive}(b)) there is no purely upstream solution (no solution for $\theta_\B{k}=\pi$) but two purely downstream solutions ($\theta_\B{k}=0)$ with different but aligned velocities. In the asymptotic limit $\bar{k}_z,\bar{k}_x\rightarrow \infty $, the hyperbola~\eqref{eq:disp_0} degenerates into the straight lines $\bar{k}_x=\pm (\bar{k}_z-\chi)/\sqrt{|\alpha|}$. In this limit the group velocity angle is maximal. Its slope is $\pm\sqrt{|\alpha|}$, since it is perpendicular to the asymptote, and therefore wave propagation is restricted to the angular sector delimited by the angles $\theta_\s=\arctan(\pm\sqrt{|\alpha}|)$.

\section{Impedance Relation and Diagram}\label{sec:impedance}

In contrast to the dispersion relation, the impedance relation depends on the polarization of the wave, as does the Poynting vector.

For the p-polarization, the impedance relation is obtained by substituting \eqref{eq:e0xz} into \eqref{eq:hoy}, so as to eliminate the wavevector quantities. This yields
\begin{subequations}\label{eq:imp_p}
\begin{equation}\label{eq:imp_p_fields}
 \frac{E_{0x}^2}{H_{0y}^2}+\frac{1}{\alpha}\frac{E_{0z}^2}{H_{0y}^2}=\frac{\mu'}{\epsilon'}.
\end{equation}
This relation may be rewritten in terms of the wave impedance components $\eta_{\tx{p}z}=E_{0x}/H_{0y}$, $\eta_{\tx{p}x}=-E_{0z}/H_{0y}$ and $\eta'=|\EO'|/|\HO'|=\sqrt{\mu'/\epsilon'}$ as
\begin{equation}\label{eq:imp_cart_p}
\alpha\eta_{\tx{p}z}^2+\eta_{\tx{p}x}^2=\alpha\eta'^2,
\end{equation}
or, alternatively, in polar form upon substituting $E_{0x}/H_{0y}=|\B{E}_0|\cos\theta_\s/|\B{H}_0|$, $E_{0z}/H_{0y}=|\B{E}_0|\sin\theta_\s/|\B{H}_0|$ (see geometrical decomposition in Fig.~\ref{fig:waves}(a)), and $\eta=|\B{E}_0|/|\B{H}_0|$ in~\eqref{eq:imp_p_fields}, as
\begin{equation}\label{eq:imp_polar_p}
\eta_\tx{p}^2\left(\alpha\cos^2\theta_\s+\sin^2\theta_\s\right)=\eta'^2\alpha.
\end{equation}
\end{subequations}
The relevant angle for the impedance diagram, given by this relation, is thus the Poynting vector angle, whereas the relevant angle for the dispersion relation, given by~\eqref{eq:neff}, was the wave vector angle, consistently with the field configuration in Fig.~\ref{fig:waves}(a). The relations~\eqref{eq:imp_p} are the equations of an ellipse for $\alpha>0$ (subluminal regime) and of a hyperbola for $\alpha<0$ (interluminal regime), both centered at $(\eta_{\tx{p}z},\eta_{\tx{p}x})=(0,0)$. 

For the s-polarization, the impedance relation is similarly obtained from~\eqref{eq:s_max} as
\begin{subequations}\label{eq:imp_s}
\begin{equation}\label{eq:imp_deriv_s}
 \frac{H_{0x}^2}{E_{0y}^2}+\frac{1}{\alpha}\frac{H_{0z}^2}{E_{0y}^2}=\frac{\epsilon'}{\mu'}.
\end{equation}
This relation may be best rewritten in terms of the wave admittance (as opposed to impedance), with the admittance components $\eta_{\tx{s}z}^{-2}=H^2_{0x}/E^2_{0y}$ and $\eta_{\tx{s}x}^{-2}=H^2_{0z}/E^2_{0y}$, as
\begin{equation}\label{eq:imp_rect_s}
 \alpha\eta_{\tx{s}z}^{-2}+\eta_{\tx{s}x}^{-2}=\alpha\eta'^{-2},
\end{equation}
or, in polar form, as
\begin{equation}\label{eq:imp_polar_s}
 \eta_\tx{s}^{-2}\left(\alpha\cos^2\theta_\s+\sin^2\theta_\s\right)=\eta'^{-2}\alpha,
 \end{equation}
\end{subequations}
which also involves the Poynting vector angle instead of the wave vector angle. The admittance relations~\eqref{eq:imp_s} are the equations of an ellipse for $\alpha>0$ (subluminal regime) and of a hyperbola for $\alpha<0$ (interluminal regime), both centered at $(\eta^{-1}_{\tx{s}z},\eta^{-1}_{\tx{s}x})=(0,0)$.

The following notes are here in order. First, the p- and s-polarization impedances, resp. given in~\eqref{eq:imp_p} and \eqref{eq:imp_s}, are inversely proportional to each other, specifically $\eta_\tx{p}/\eta'=\eta'/\eta_\tx{s}$. Second, their components along the direction of motion found from~\eqref{eq:e0xz} and~\eqref{eq:h0xz} as
\begin{equation}\label{eq:trans_imp}
    \eta_{\tx{p}z}=\frac{E_{0x}}{H_{0y}}=\frac{k_z-\omega\chi/c}{\omega\epsilon'\alpha}, \quad \eta_{\tx{s}z}=\frac{E_{0y}}{H_{0x}}=-\frac{\omega \mu' \alpha}{k_z-\omega\chi/c}
    \end{equation}
are in agreement  with the transverse impedance expression in~\cite{radi2015,tretyakov1994}.

\begin{figure}[ht] \includegraphics[width=1\columnwidth]{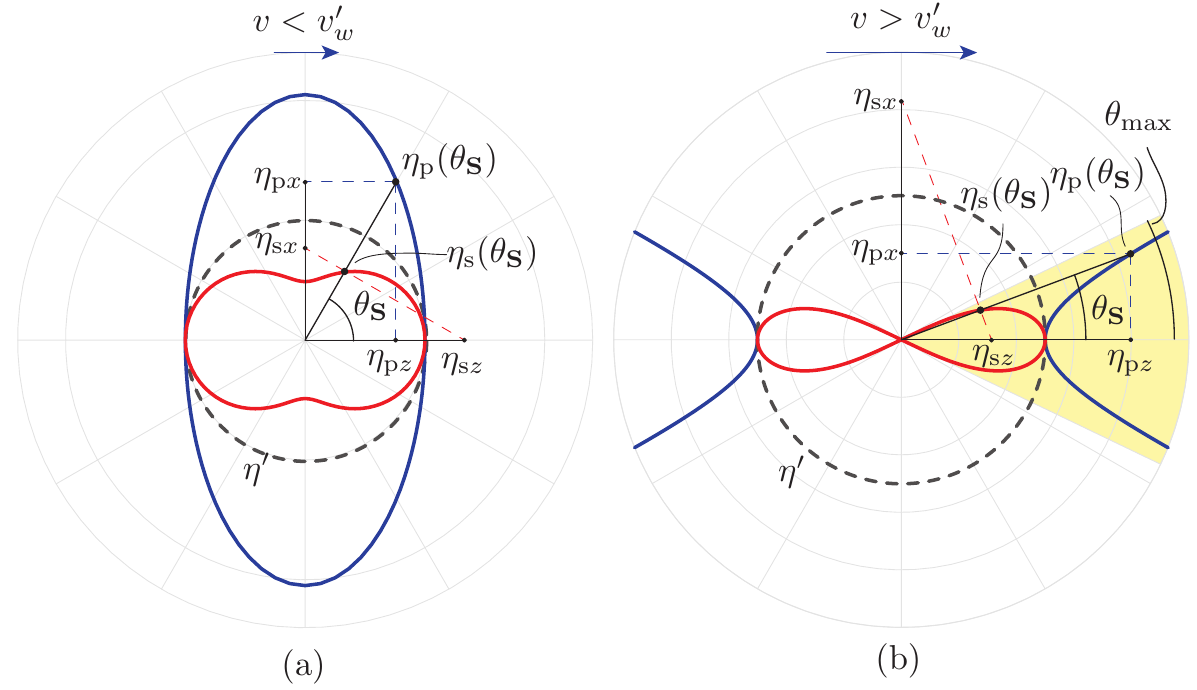}
\caption{Impedance diagram [Eq.~\eqref{eq:imp_p}),~\eqref{eq:imp_s}] for a moving dielectric (fluid) medium with $n'=2$ ($v_\text{w}'=c/2$) and $\eta'=\eta_0/2$. The dashed circles correspond to the isotropic curves at $v=0$. (a)~Subluminal regime, with velocity $v=0.45c<v_\text{w}'$. (b)~Interluminal regime, with $v=0.8c>v_\text{w}'$. The highlighted region corresponds to the region of physical solutions, with $\theta_\tx{max}=\arctan{\pm\sqrt{|\alpha}|}$.}\label{fig:imp}
\centering
\end{figure}

Figure~\ref{fig:imp} plots typical impedance diagrams. Figure~\ref{fig:imp}(a) corresponds to the subluminal regime. The p-polarization impedance curve is an ellipse, while the s-polarization impedance curve is the inverse of an ellipse, an elliptic limaçon. The figure shows that the impedance is unaffected by motion for pure-downstream and pure-upstream propagation, i.e., $\eta_{\tx{s,p}}(0^\circ)=\eta_\tx{s,p}(180^\circ)=\eta'$ \footnote{Indeed, the transverse components of the $\B{E}$ and $\B{H}$ fields transform the same way under Lorentz transformation and their ratio is thus invariant.}, while for all the other angles $\eta_\tx{p}>\eta'$ and $\eta_\tx{s}<\eta'$. Note that the p-polarization impedance $x-$ and $z-$ components are found by simple orthogonal projections, whereas the s-polarization $x-$ and $z-$ impedance components, which are mathematically expressed as $\eta_{\tx{s} x}=\eta_\tx{s}/ \sin\theta_\s$ and $\eta_{\tx{s} z}=\eta_\tx{s}/ \cos\theta_\s$, as seen by comparing~\eqref{eq:imp_rect_s} to~\eqref{eq:imp_polar_s}, are not found using orthogonal projections, but are rather found as illustrated in Fig.~\ref{fig:imp}. The opposite would be true if we had chosen an admittance representation, and the impedance/admittance representations are naturally best suited to the p/s-polarizations, but we wanted to show the two polarization impedances on the same diagram in order to compare them. 

Figure~\ref{fig:imp}(b) corresponds to the interluminal regime. The p-polarization impedance curve is now a hyperbola, and the s-polarization impedance curve is a hyperbolic limaçon. The same remarks apply: the impedances for pure-downstream and pure-upstream propagation are unaffected by the motion, and $\eta_\tx{p}>\eta'$ and $\eta_\tx{s}<\eta'$ for all other angles . As was shown in Sec.~\ref{sec:refr_ind_vel} (Fig.~\ref{fig:refractive}(b)), the Poynting vector is restricted to the angular sector $\theta_\s\leq\arctan(\pm\sqrt{|\alpha}|)$. Since the impedance is only meaningful in this sector, the solutions in the two left quadrants do not correspond to physical solutions.

\section{Scattering at the Stationary Interface}\label{sec:scatter}

We shall now study the scattering of waves at a stationary interface between free space and the moving medium (Fig.~\ref{fig:Fizeau}). The interface is positioned in the $xy$ plane at $z=0$, the medium propagates in the $z$ direction, and hence perpendicularly to the interface, and the incident wave is in the $xz$-plane. Our purpose is to determine what waves are scattered across the interface and under which angles such scattering occurs in terms of both phase and group velocities. 

The incident, reflected and transmitted fields, all of the form~\eqref{eq:th_fields}, must satisfy at the interface ($z=0$) the spatial and temporal phase matching conditions
\begin{subequations}\label{eq:phase_match}
\begin{equation}\label{eq:phase_match_kx}
k_{x\tx{i}}=k_{x\tx{r}}=k_{x\tx{t}},
\end{equation}
\begin{equation}\label{eq:phase_match_omega}
\omega_\tx{i}=\omega_\tx{r}=\omega_\tx{t}.
\end{equation}
\end{subequations}
The first equalities in these relations immediately indicate that the reflection angle must always be equal to the incidence angle, $\theta_\text{r}=\theta_\text{i}$, i.e., that the reflection must always be specular. To determine the transmission phase angle, $\theta_{\mathbf{k}\tx{t}}$, and the transmission group angle, $\theta_{\mathbf{S}\tx{t}}$, which are generally given  by~\eqref{eq:thetak} and~\eqref{eq:thetak}, respectively, one must first express the corresponding quantities $k_{x\tx{t}}$, $k_{z\tx{t}}$ and $\omega$ as a function of the incident wave quantities so as to satisfy the other conditions in~\eqref{eq:phase_match}. We have already from these relations $k_{x\text{t}}=k_{x\text{i}}$ and $\omega_{\text{t}}=\omega_{\text{i}}$. We solve then~\eqref{eq:disp_0} for $k_z$, which yields $k_{z\tx{t}}=\chi\omega_\tx{t}/c\pm\sqrt{\alpha^2{n_2'}^2\omega_\tx{t}^2/c^2-\alpha k_{x \tx{t}}^2}$, where we apply $k_{x\text{t}}=k_{x\text{i}}$ and $\omega_{\text{t}}=\omega_{\text{i}}$. Substituting these results into~\eqref{eq:angles_pg} finally yields the sought-after angles in terms of the angle and frequency of the incident wave
\begin{subequations}\label{eq:angles_pgi}
\begin{equation}\label{eq:thetaki}
\theta_{\K\text{t}}(\theta_{\text{i}},\omega_\text{i})
=\arctan\left(\frac{k_\text{i}\sin\theta_{\text{i}}}{\chi\frac{\omega_\tx{i}}{c}\pm\sqrt{\alpha^2{n_2'}^2\omega_\tx{i}^2/c^2-\alpha k_\text{i}^2\sin^2\theta_{\text{i}}}}\right),
\end{equation}
\begin{equation}\label{eq:thetasi}
\theta_{\s\text{t}}(\theta_{\text{i}},\omega_\text{i})
=\arctan\left(\frac{\pm\alpha k_\text{i}\sin\theta_{\text{i}}}{\sqrt{\alpha^2{n_2'}^2\omega_\tx{i}^2/c^2-\alpha  k_\text{i}^2\sin^2\theta_{\text{i}}}}\right),
\end{equation}
\end{subequations}
where the dependence on the medium motion velocity ($v$) is embedded in the parameters $\alpha$ and $\chi$, given in~\eqref{eq:alpha_chi}.

\begin{figure}[h!] \includegraphics[width=1\columnwidth]{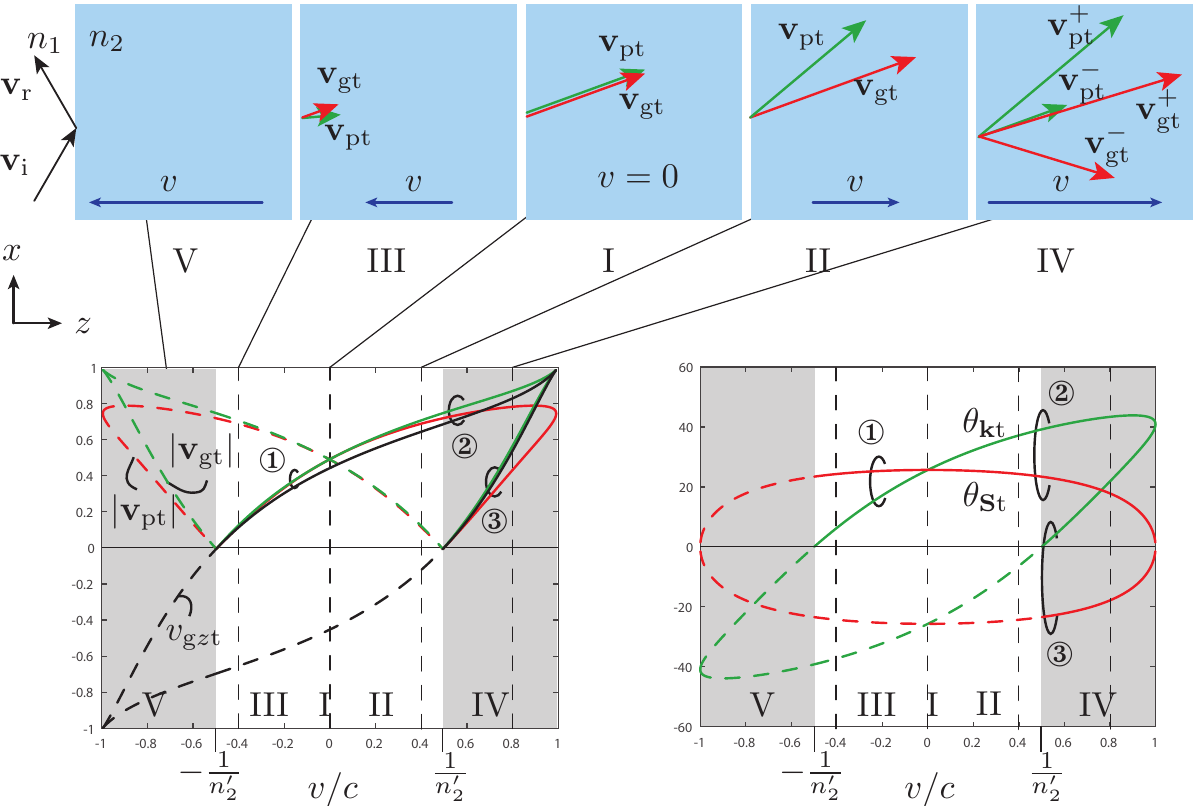}
\centering
\caption{Wave scattering at the interface in Fig.~\ref{fig:Fizeau} for the parameters $\theta_\tx{i}=60^\circ$, $n_1'=1$, and $n_2'=2$. Bottom-left panel: group velocity magnitude, phase velocity magnitude, and $z$-component of the group velocity, versus medium velocity. Bottom-right panel: group velocity and phase velocity angles versus medium velocity. The numbers \ding{172}, \ding{173} and \ding{174} indicate the segments of physical solutions, corresponding to $v_{\text{g}z}>0$, while the dashed lines indicate nonphysical solutions. Top panels: ray configurations corresponding to the five regions indicated in the other two figures (incident and reflected rays indicated only in the left-most panel.).}
\label{fig:angles}
\end{figure}

Figure~\ref{fig:angles} plots the moving-medium phase and group velocities, resp. given by~\eqref{eq:vp} and~\eqref{eq:vg}, the corresponding scattering (transmission) angles at the interface in Fig.~\ref{fig:Fizeau}, resp. given by~\eqref{eq:thetaki} and~\eqref{eq:thetaki}, and the related rays. One distinguishes five main regimes of modulation velocity ($v$), corresponding to the regions labeled I to V in the figure. 

Region~I ($v=0$) corresponds to the stationary regime, where wave scattering reduces the usual Snell reflection and refraction, and where $\mathbf{v}_\text{pt}=\mathbf{v}_\text{gt}$ since the medium is isotropic at rest. 

Region~II ($0<v<c/n_2'$) corresponds to the subluminal co-directional ($\sgn(v)=\sgn(v_\text{pz})$) regime, where the Fresnel-Fizeau drag boosts both $\mathbf{v}_\text{pt}$ and $\mathbf{v}_\text{gt}$ (swimmer progressing faster downstream), and where we observe a splitting between the directions of the two velocities. Only the mathematical solution branch with $v_{\text{g}z\text{t}}>0$ is physical, since the wave is launched towards positive $z$.

Region~III ($c/n_2'<v<c$) corresponds to the subluminal contra-directional ($\sgn(v)=-\sgn(v_\text{pz})$) regime, where the Fresnel-Fizeau drag effect is now negative (swimmer progressing slower upstream). Again, only the solution with positive $v_{\text{g}z\text{t}}$ is physical, and we observe a flipping between the directions of the phase and group velocities compared to the subluminal co-directional case.

Region~IV ($c/n_2'<v$) corresponds to the superluminal co-directional regime, where the Fresnel-Fizeau boost is, unsurprisingly, even greater than in the co-directional subluminal regime. The two mathematical solutions have a positive $v_{\text{g}z\text{t}}$, and are therefore both physical. Note that the two group velocities are direction-wise symmetric but of different magnitudes while the two phase velocities differ in both direction and magnitude in this regime.

Finally, Region~V ($v<-c/n_2'$) corresponds to the superluminal contra-directional regime, where no wave is transmitted into the medium, because the wave reaching the interface is pushed back by the faster contra-directional drag, consistently with the fact that $v_{\text{g}z\text{t}}<0$ for the two mathematical solution branches.

Let us now present, alternatively to the mathematical formulas used above, a graphical method that offers a quick qualitative resolution of the problem as well as deep physical insight into it~\cite{deck2018,deck2019}.

\begin{figure}[ht] \includegraphics[width=1\columnwidth]{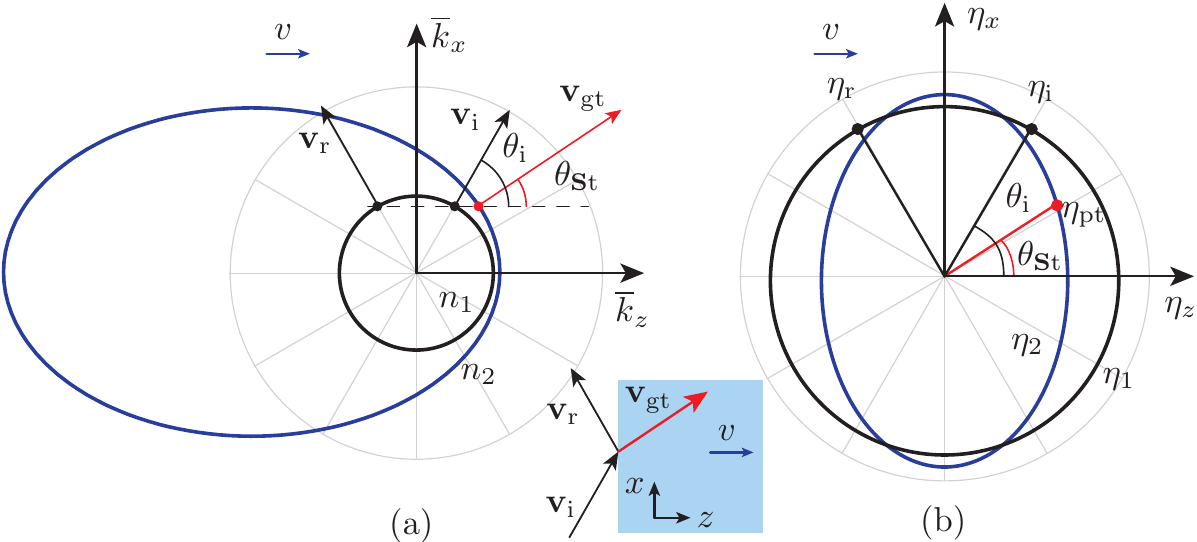}
\centering
\caption{Graphical solution for the subliminal co-directional regime, for p-polarization, with $v=0.6c$ and $\theta_\tx{i}=60^\circ$. (a)~Refractive index diagram. Inset: scattering in space. (b)~Impedance diagram.}
\label{fig:scat1}
\end{figure}

Figure~\ref{fig:scat1} presents the graphical solution for the subluminal co-directional regime. The procedure is as follows: 1)~plot the refractive index relations and the impedance relations for the media, i.e., $n_1(\theta_\K)$ and $n_2(\theta_\K)$, and $\eta_1(\theta_\s)$ and $\eta_2(\theta_\s)$, in corresponding diagrams; 2)~locate the incident wave on the $n_1$ curve for the selected $\theta_{\K \tx{i}}$; 3)~trace a horizontal line passing through this point, corresponding to the conservation of $k_x/k_0$ associated with the phase matching condition, and identify the intersections of this line with the refractive index curves; 4)~at the intersection points, trace the Poynting vector angles $\theta_{\s\tx{r}}$ and $\theta_{\s\tx{t}}$ as the normal directions to the curves; 5)~report these angles on the impedance diagram to locate the solutions $\eta_\tx{i,r}=\eta_1(\theta_{\s\tx{i,r}})$ and $\eta_\tx{t}=\eta_2(\theta_{\s\tx{t}})$.

\begin{figure}[ht] \includegraphics[width=1\columnwidth]{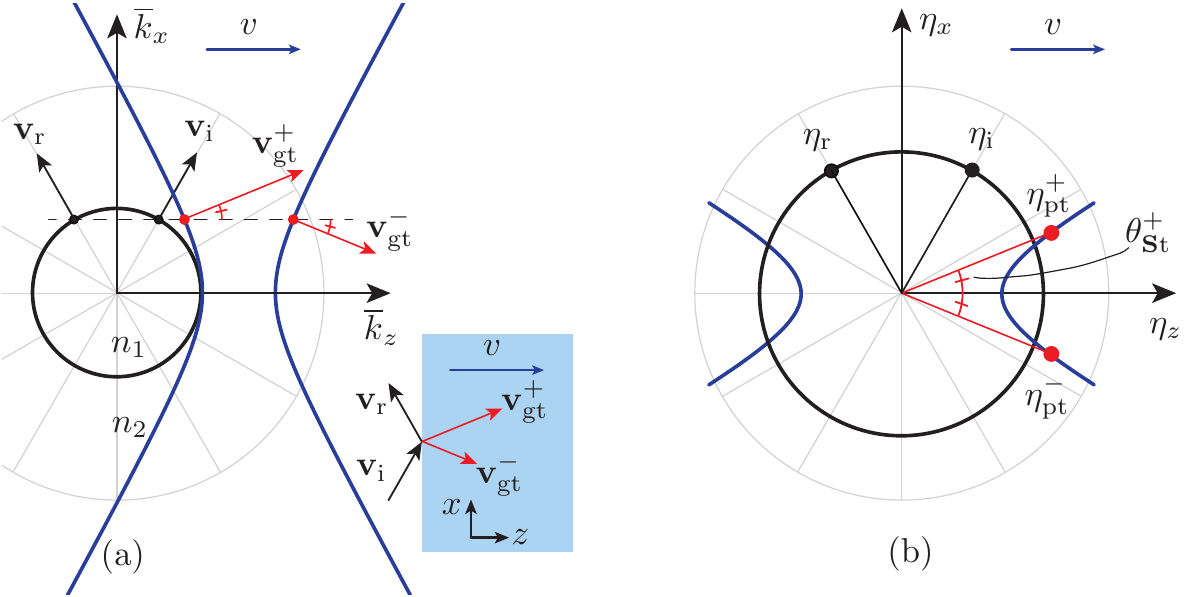}
\centering
\caption{Same as Fig.~\ref{fig:scat1}, but for the interluminal regime, with $v=0.9c$.}
\label{fig:scat2}
\end{figure}

Figure~\ref{fig:scat2} presents the graphical solution for the interluminal co-directional regime, which follows the same procedure. From Fig.~\ref{fig:scat2}(a), we confirm that the group velocities of the two waves in the moving medium propagate symmetrically with respect to the $z$ axis, consistently with Fig.~\ref{fig:angles}. In Fig.~\ref{fig:scat2}(b), we report the angles found in Fig.~\ref{fig:scat2}(a) onto the impedance diagram and conclude, from the symmetry of the impedance curve, that the two  transmitted waves see the same impedance, i.e., $\eta_\tx{t}^+=\eta_\tx{t}^-$.

We have plotted the diagrams only for the p-polarization, but the same method can be applied for the s-polarization.

We have determined so far scattered waves in terms of phase and group velocity vectors. To complete the resolution of the scattering problem, we still have to determine the scattering coefficients. For the subluminal regime, the reflection and transmission coefficients are easily found by enforcing the continuity of the tangential electric and magnetic fields at the interface~\cite{censor1969}, resp. given by~\eqref{eq:p_max} and~\eqref{eq:s_max} for the p- and s-polarizations, as
\begin{equation}
\Gamma_\tx{p}=\frac{H_{y\tx{r}}}{H_{y\tx{i}}}=\frac{\eta_{\tx{p}z1}-\eta_{\tx{p}z2}}
{\eta_{\tx{p}z1}+\eta_{\tx{p}z2}},\quad T_\tx{p}=\frac{H_{y\tx{t}}}{H_{y\tx{i}}}=\frac{2\eta_{\tx{p}z1}}{\eta_{\tx{p}z1}+\eta_{\tx{p}z2}},
\end{equation}
\begin{equation}
\Gamma_\tx{s}=\frac{E_{y\tx{r}}}{E_{y\tx{i}}}=\frac{\eta_{\tx{s}z1}^{-1}-\eta_{\tx{s}z2}^{-1}}
{\eta_{\tx{s}z1}^{-1}+\eta_{\tx{s}z2}^{-1}},\quad T_\tx{s}=\frac{E_{y\tx{t}}}{E_{y\tx{i}}}=\frac{2\eta_{\tx{s}z1}^{-1}}{\eta_{\tx{s}z1}^{-1}+\eta_{\tx{s}z2}^{-1}}.
\end{equation}
where the transverse impedances, defined in~\eqref{eq:trans_imp}, can be found geometrically by performing the projections described in Sec.~\ref{sec:impedance}. In contrast, for the interluminal regime, it is not easy to calculate the scattering coefficients. We recall that either no wave (contra-directional case) or two waves (co-directional case) are transmitted, leading to either only a reflected wave or one reflected and two transmitted waves, for upstream or downstream propagation, respectively. This respectively leads to an overdetermined set of equations (one unknown and two continuity conditions) or an underdetermined one (three unknowns for 2 continuity conditions). We defer the resolution of this problem to an ulterior publication.

\section{Generalization to Oblique Motion}

\begin{figure}[ht] \includegraphics[width=0.5\columnwidth]{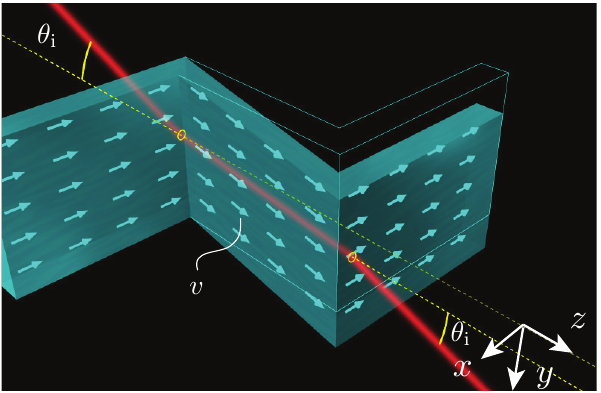}
\centering
\caption{Same as Fig.~\ref{fig:Fizeau}, but sheared along the $xz$ plane so as to have the medium moving at an angle with respect to the interface.}
\label{fig:fizeau_slant}
\end{figure}

So far, the medium motion was directed perpendicular to the interface. We now generalize the graphical method described in Sec.~\ref{sec:scatter} to arbitrary angles. This corresponds to a shearing of the center portion of the apparatus of Fig.~\ref{fig:Fizeau}, as illustrated in Fig.~\ref{fig:fizeau_slant}.

\begin{figure}[ht] \includegraphics[width=1\columnwidth]{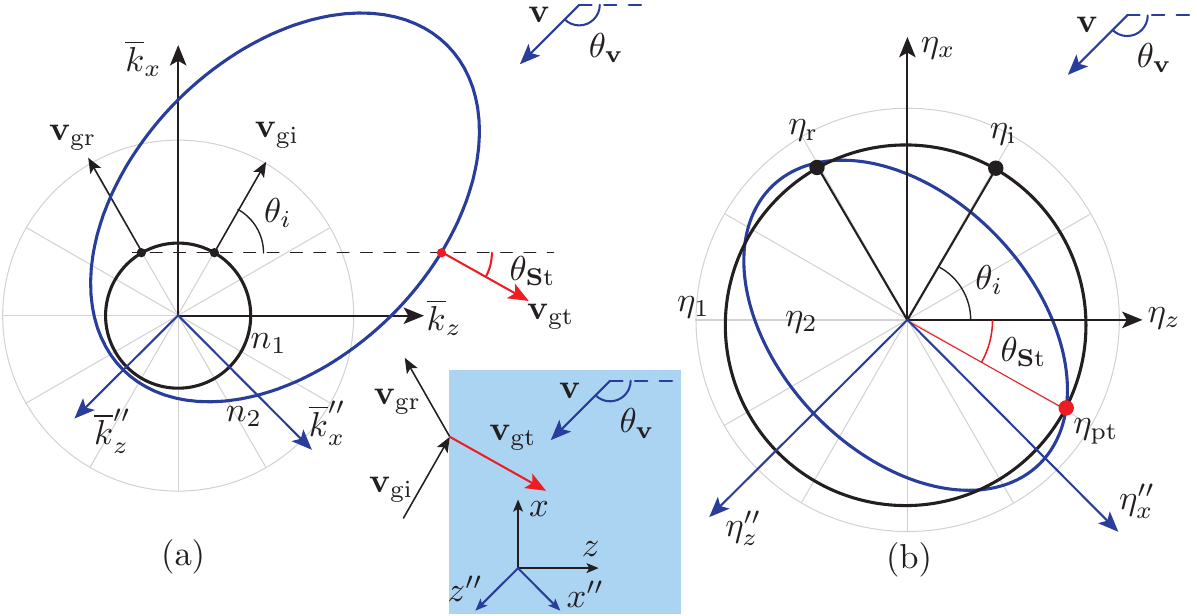}
\centering
\caption{Same as Fig.~\ref{fig:scat1}, but for medium motion at angle $\theta_\B v=135^\circ$.}
\label{fig:scat3}
\end{figure}

Figure~\ref{fig:scat3} presents the graphical solution for a subluminal contra-propagating problem with velocity forming a $135\deg$ angle with respect to the $z$ direction. To account for the angle of the medium velocity, the dispersion and impedance patterns of the bulk medium, with coordinates of the bulk medium, $k_x''$, $k_z''$ and $\eta_x'', \eta_z''$, where the double primes are used to avoid confusion with the primes of the Lorentz transformation, are rotated by $135\deg$ with respect to the coordinates of the interface. Once the isofrequency and impedance patterns have been drawn, the usual boundary condition, $k_{x\tx{i}}=k_{x\tx{r}}=k_{x\tx{t}}$, is applied, the interface being still positioned at $z=0$. It is interesting to note that the oblique motion gives rise to negative refraction, which was not possible for perpendicular motion. Note that negative refraction for a medium moving at a subluminal velocity parallel to the interface, corresponding to a rotation of $\theta_\B v=90\deg$, was reported in~\cite{grzergorczyk2006}. The rotation in this scenario also increases the wave impedance.

\section{Connection with Spacetime-Modulated Systems}\label{sec:con_ST_syst}

\begin{figure}[ht] \includegraphics[width=1\columnwidth]{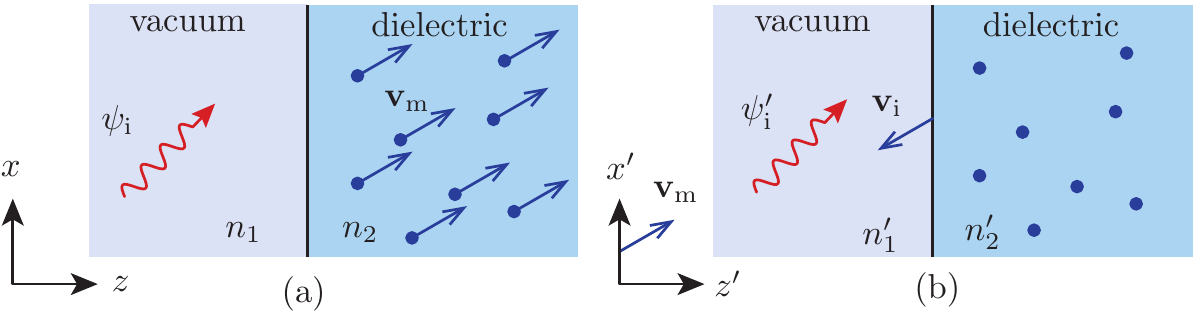}
	\caption{Comparison of two closely related problems involving motion. (a)~Problem studied in this paper. (b)~Related spacetime-modulated problem. In both cases, we assume that the incident medium is vacuum, i.e., $n_1=n_1'=1$, and the transmission medium a dielectric fluid that is isotropic at rest.}
	\label{fig:spacetime}
\end{figure}

The problem studied until this point is closely related to the problem of spacetime modulation (as opposed to motion) where the medium parameters are space- and time-invariant and the interface is moving in the form of a perturbation induced by an external modulation, without involving any transfer of matter~\cite{caloz2020,caloz2020_2}. Figure~\ref{fig:spacetime} compares the two problems, with Fig.~\ref{fig:spacetime}(a) representing the problem studied here and Fig.~\ref{fig:spacetime}(b) the related spacetime-modulated problem. In the former case, the reference frame is the rest frame (vacuum), corresponding to the $xz$ coordinate system, where the interface is stationary and the particles of the medium (e.g., fluid molecules) move in the direction $\B v_\tx{m}$. In the latter case, the reference frame is the moving frame (dielectric), corresponding to the $x'z'$ coordinate system, where the particles of the medium (e.g., fluid molecules) are stationary, while the interface moves in the direction $\B v_\tx{int}=-\B v_\tx{m}$ and .

The two problems are equivalent by constitution and by transformation but different in terms of the physics experienced by a wave incident on the interface: 1)~Constitutional equivalence -- In both cases, we have the same set of particles and the same interface between vacuum and the medium formed by these particles, the difference being only a change of system of coordinates; 2)~Transformation equivalence -- The two problems are Lorentz transformations of each other, and each is therefore perfectly determined by the other; 3)~Physics difference\footnote{The simplest example would be the case where the dielectric reduces to a single charged particle, with charge $q$ (or, more generally, a plasma). The problem of Fig.~\ref{fig:spacetime}(b) is then a problem of electrostatics, since $\partial q/\partial t=0$, whereas the problem of Fig.~\ref{fig:spacetime}(a) is then a problem of electrodynamics and electromagnetics,  since $\partial q/\partial t\neq 0=i$.} -- in the case of Fig.~\ref{fig:spacetime}(a), the wave incident from vacuum sees a stationary interface followed by a bianisotropic medium, with electric-magnetic coupling being produced by the motion of particles~\cite{minkowski1908,kong2005}, while in the case of Fig.~\ref{fig:spacetime}(a) it sees an isotropic medium across a moving interface, so that distinct scattering phenomenologies occur in the two cases~\cite{caloz2020,caloz2020_2}, with the most striking difference being the presence of Doppler frequency shifting in the latter case due to the interface motion and not in the former due to phase matching (Eq.~\eqref{eq:phase_match_omega}).

\section{Conclusion}

We have derived the refractive index and impedance relations for bulk moving media and plotted the results in related diagrams in both the subluminal and interluminal regimes. Moreover, we have mathematically and graphically determined the phase and group velocities of the waves scattered at a stationary interface bounding a moving medium, for medium motion perpendicular and oblique with respect to the interface. Finally, we have discussed a connection between this problem and the problem of a spacetime modulated medium. 

The experimental implementation of the problem solved in the paper seems most challenging at the high end of the the subluminal regimes and in the interluminal regimes, given the relativistic medium velocities involved there, even if one would resort to the latest fluid dynamics technologies. However, spacetime modulations of relativistic velocities are perfectly attainable in practice, and are related to the problem studied here by simple Lorentz transformations. The results presented here may therefore be of interest for the practically-realizable currently emerging spacetime modulated metamaterials~\cite{caloz2020,caloz2020_2}.
\vspace{6pt} 



\authorcontributions{Conceptualization, Z.D and C.C; methodology, Z.D and C.C.; software, Z.D.; validation, Z.D, X.Z. and C.C; formal analysis, Z.D, X.Z. and C.C.; investigation, Z.D.; resources, C.C.; data curation, Z.D; writing---original draft preparation, Z.D and C.C.; writing---review and editing, Z.D and C.C; visualization, Z.D. and C.C.; supervision, C.C.; project administration, C.C.; funding acquisition, C.C. All authors have read and agreed to the published version of the manuscript.}

\conflictsofinterest{The authors declare no conflict of interest.} 

\end{paracol}
\reftitle{References}


\externalbibliography{yes}
\bibliography{moving_medium}

\end{document}